\documentclass[fleqn,usenatbib]{mnras}

\usepackage{newtxtext,newtxmath}

\usepackage[T1]{fontenc}

\DeclareRobustCommand{\VAN}[3]{#2}
\let\VANthebibliography\thebibliography
\def\thebibliography{\DeclareRobustCommand{\VAN}[3]{##3}\VANthebibliography}

\usepackage{graphicx}	
\usepackage{amsmath}	
\usepackage{comment}

\newcommand{\ppdot}{$P$--$\dot{P}$}

\title[TPA Single Pulses]{The Thousand-Pulsar-Array programme on MeerKAT XIX: Single-pulse data analysis, nulling and pulse energy distributions}

\author[M.~J.~Keith et al.]{Michael J. Keith,$^{1}$\thanks{E-mail: mkeith@pulsarastronomy.net}
Patrick Weltevrede$^{1}$,
Lucy Oswald$^{2,3,4}$,
Aris Karastergiou$^{4}$,
Xiaoxi Song$^{5}$,
\newauthor
Haoyue Wang$^{1}$,
Jui-An Hsu$^{1}$,
Simon Johnston$^{6}$,
Geoff Wright$^{1}$,
Matthew Bailes$^{7,8}$,
Maciej Serylak$^{9}$
\\
$^{1}$Jodrell Bank Centre for Astrophysics, Department of Physics and Astronomy, University of Manchester, Manchester M13 9PL, UK\\
$^{2}$School of Physics \& Astronomy, University of Southampton, Southampton SO17 1BJ, UK\\
$^{3}$Astrophysics Group, Cavendish Laboratory, University of Cambridge, J. J. Thomson Avenue, Cambridge CB3 0HE, UK\\
$^{4}$Department of Astrophysics, University of Oxford, Denys Wilkinson Building, Keble Road, Oxford OX1 3RH, UK\\
$^{5}$ASTRON, the Netherlands Institute for Radio Astronomy, Oude Hoogeveensedijk 4,7991 PD Dwingeloo, The Netherlands \\
$^{6}$Australia Telescope National Facility, CSIRO, Space and Astronomy, PO Box 76, Epping, NSW 1710, Australia\\
$^{7}$Centre for Astrophysics and Supercomputing, Swinburne University of Technology, PO Box 218, Hawthorn, VIC 3122, Australia\\
$^{8}$OzGrav: The ARC Centre of Excellence for Gravitational Wave Discovery, Hawthorn, VIC 3122, Australia\\
$^{9}$SKA Observatory, Jodrell Bank, Lower Withington, Macclesfield SK11 9FT, UK
}

\date{Accepted XXX. Received YYY; in original form ZZZ}

\pubyear{\the\year{}}

\begin{document}
\label{firstpage}
\pagerange{\pageref{firstpage}--\pageref{lastpage}}
\maketitle

\begin{abstract}
We present the Thousand Pulsar Array (TPA) single-pulse data set, obtained with the MeerKAT radio telescope and comprising time-series observations of 1192 pulsars, typically containing $\sim 1000$ consecutive pulses per source. We describe the MeerTime Single Pulse software pipeline which calibrates the data and automatically excises interference signals to produce data products suitable for typical single-pulse studies.
To demonstrate the capabilities of the dataset, we carry out a population-level study of phase-averaged single-pulse energy distributions and nulling behaviour.
Pulse energy distributions are modelled within a Bayesian framework choosing from a range of intrinsic energy distributions, and including an explicit nulling fraction. 
We find that approximately half of the pulsars require multi-component intrinsic energy distributions, while the remainder are consistent with single-component models. Nulling is detected or constrained for most pulsars in the sample, and both the occurrence and inferred nulling fraction show systematic variation across the \ppdot\ diagram. In particular, nulling fractions increase with spin period and exhibit only a weak dependence on period derivative.
We also examine trends in the preferred forms of pulse energy distributions as a function of spin-down luminosity, finding modest evidence for population-level evolution. Estimates of single-pulse luminosities indicate that individual pulses can exceed the long-term average luminosity by large factors, particularly for low-$\dot{E}$ pulsars. These results characterise the statistical properties of single-pulse emission across a large pulsar sample and highlight the limitations of phase-averaged energy distributions for capturing the full complexity of pulsar emission variability.
\end{abstract}

\begin{keywords}
pulsars: general
\end{keywords}



\section{Introduction}

Although the time-averaged shape of pulses from radio pulsars is usually highly stable, individual pulses can vary considerably from pulse to pulse. Pulse-to-pulse variability of pulsars comes in many forms, including nulling \citep{1970Natur.228...42B}, mode-changing \citep[e.g.][]{Wen2020}, variation in pulse intensity \citep[e.g.][]{smith73}, random phase jitter of sub-pulses \citep[e.g.][]{pbs+21, khb+2024}, or organised drifting or modulation of sub-pulses \citep[e.g.][]{Weltevrede2007, Song2023}. The time-variability from radio pulsars is important to investigate and understand, both because doing so directly probes the physics of the neutron star magnetosphere \citep[e.g.][]{Song2023}, and because such variability presents a problem for precision pulsar timing for tests of gravitation and searches for gravitational waves \citep[e.g.][]{jcc+2024}.

Previous studies of pulsar pulse energy distributions have suggested that typical pulsar emission energies are drawn from a log-normal distribution (e.g. \citealp{cjd01,cjd04}), although \cite{Mickaliger2018} suggested that a power law divided by an exponential leads to a better fit for most of their survey of 264 pulsars and 14 Rotating Radio Transients, and showed that some pulsars deviate from a log-normal power distribution due to excess high-energy pulses. \cite{bjb+12} found that, for 315 pulsars detected in the mid-latitude portion of the Parkes High Time Resolution Universe survey, about 40 per cent showed a log-normal single-pulse energy distribution and the rest showed behaviour that was characterised as either Gaussian or erratic. Separately, it seems that giant pulses tend to follow a power-law distribution \citep{lundgren_1995,2002MNRAS.332..109J}, and it is suggested that giant pulses originate not from near the polar cap, like other radio pulses, but in or close to the current sheet instead \citep{Lyubarsky2019,Philippov2019}.

Nulling in pulsars is seen over a wide range of timescales, with each null lasting typically fewer than 5 pulses \citep{basu17}, but long term nulling can be seen where emission ceases for months or years \citep[e.g.][]{klo+96}. Nulls are often seen to have a characteristic timescale or quasi-periodic behaviour for a given pulsar, which is typically different to that of the sub-pulse modulation or drifting, indicating a different origin to the nulling \citep{basu17}. Nulling is observed to be a broad-band phenomenon \citep{Weltevrede2006,Weltevrede2007, gjk+14, Naidu2017} but can also appear only in selective frequency bands \citep{Bhat2007}. It is not yet clear whether nulling is an extreme form of mode-changing or an independent phenomenon \citep[e.g.][]{Naidu2017, 2018MNRAS.475.2375N}.

Recent years have seen substantial progress in studies of pulsar nulling driven by the availability of sensitive single-pulse observations and increasingly sophisticated statistical methodologies. \citet{ksfv18} introduced a Gaussian mixture modelling framework for estimating nulling fractions directly from pulse-energy distributions, allowing null and burst states to be statistically separated without requiring hard thresholds. This method was also adapted and expanded by \citet{brook26} in a study of time evolution of nulling fractions in a sample of pulsars. \citet{Grover2026} analysed nulling pulsars using long-duration Murchison Widefield Array observations at 140--170\,MHz, introducing the ``$N_{\rm sum}$'' method in which randomly selected groups of pulses are summed in order to statistically separate weak burst emission from the noise floor. The large number of pulses available per observation enabled detailed studies of  nulling behaviour and quasiperiodic nulling phenomena.

Although there are many detailed single-pulse studies of bright individual pulsars, comparatively few studies have characterised pulse-to-pulse emission across large pulsar samples. Single-pulse studies of pulsars are inherently limited in integration time to a fraction of the pulse width, and therefore signal-to-noise ratio (S/N) is dominated by the instantaneous sensitivity of the telescope. This means that it is difficult to collect a sufficient quantity of high S/N data for a population of sources. However, the large collecting area and wide bandwidth of MeerKAT are particularly beneficial for single-pulse studies of pulsars. As a result, a key goal of the Thousand Pulsar Array (TPA) programme on the MeerKAT radio telescope was to produce a legacy dataset of $\sim 1000$ individual pulses from each of $\sim 1000$ pulsars \citep{TPA_1,TPA_2}. 

The most recent large-scale surveys of pulse energy distributions were carried out with Murriyang, the Parkes radio telescope: that of 315 pulsars presented by \citet{bjb+12}, observed as part of the Parkes High Time Resolution Universe survey, and that of 264 pulsars presented by \cite{Mickaliger2018} as part of the Parkes Multibeam Pulsar Survey. With six times the sensitivity and four times the number of pulsars, the TPA single-pulse census represents a marked improvement in the overall number of pulsars for which high signal-to-noise measurements of individual pulsars have been made. Unlike survey-based datasets which typically have a fixed integration time, the TPA single pulse census chose an integration time to capture at least 1024 pulses from most pulsars, with the exception of a small number of long-period pulsars ($2$ seconds), for which 512 pulses are recorded (see \citealt{Song2023} for an in-depth description of the sample). This provides a broadly homogeneous dataset across a large fraction of the known pulsar population for studying short-term variability, complementing existing deep studies of individual pulsars with tens of thousands of pulses from a small number of pulsars \citep[e.g.][]{whh+20}, or vast numbers of pulses from millisecond pulsars \citep{pbs+21,msb+22}.

In this paper we present the Meertime Single-pulse Pipeline (MTSP), developed specifically to process single-pulse observations from the MeerKAT telescope, and the resultant Thousand Pulsar Array single-pulse data set.
The resulting calibrated and cleaned data for the full sample of pulsars are available via Zenodo\footnote{\url{http://doi.org/10.5281/zenodo.18980771}} and via a web interface\footnote{\url{http://psrweb.jb.man.ac.uk/tpa/singlepulse}}.
The single-pulse subpulse modulation properties of the data set have previously been published by \cite{Song2023}, and the polarization properties of the pulsars with the brightest single pulses in the data set were studied by \cite{Johnston2024}. Studies of drifting sub-pulses in individual pulsars have also been published from these data \citep{2022ApJ...934...23S,2025MNRAS.538.1063H,2025MNRAS.544..234W}.
Here, we present an analysis of the pulse energy distributions of the TPA single-pulse data set. We use the large sample size to probe the evolution of nulling fraction and pulse energy distribution as a function of the pulsar's period ($P$) and period derivative ($\dot{P}$).

In Sections \ref{sec:datarecording} and \ref{sec:MTSP} we explain how the data were recorded and processed with the MTSP. We present our methodology for calculating pulse energy distributions and nulling fraction in Section \ref{sec:energydistribs}. Section \ref{sec:resultsdiscussion} is devoted to presentation of the results of these measurements, discussion of them in the wider context of pulse single-pulse variability, and the conclusions are given in Section \ref{sec:conclusions}.

\section{The Meertime time-series data recording}
\label{sec:datarecording}
The Meertime single-pulse data are recorded as channelised time-series \textsc{psrfits} files (sometimes known as ``filterbank'' or ``search-mode'' data) using the Pulsar Timing User Supplied Equipment (PTUSE) hardware.
A detailed description of the Meertime observing system can be found in \citet{2020PASA...37...28B}.
For `L-band' observations carried out as part of the TPA, the primary focus of this paper, the data are recorded with 0.8359375$\,$MHz channels, sampled every $\sim$38.28$\,\mu$s.
Observations for other Meertime sub-programmes follow the same general principles, but may be recorded with higher or lower resolution, for example observations of millisecond pulsars are typically observed using four times as many 0.208984375$\,$MHz channels sampled every $\sim$9.57$\,\mu$s.

\begin{table}
    \centering
    \caption{Time-series observing bands used by Meertime. Data in 2019 March and 2019 April are recorded in two sub-bands, which are combined in subsequent processing. The number of frequency channels ($N_\mathrm{chan}$) is given for the Thousand Pulsar Array project.}
    \label{bands_table}
    \begin{tabular}{c|ccc}
    Dates & Centre Freq & Bandwidth & $N_\mathrm{chan}$ \\
          &    (MHz) & (MHz) & \\
    \hline

2019 Mar --- 2019 Apr & 1283.58203125 & 856 & 1024 \\
2019 May --- 2020 Feb 10 & 1283.58203125 & 642 & 768 \\
        2020 Feb 10 --- onward & 1283.58203125 & 856 & 1024 \\
    \end{tabular}
    
\end{table}

The PTUSE computers split the observed data into two equal sub-bands, processed independently by one of the two GPUs in each machine.
From 2019 May onwards the recording of the two sub-bands was synchronised so that the raw time-series data could be combined during recording to produce a single \textsc{psrfits} file containing the entire observing band.
However, for data recorded prior to 2019 May the two sub-bands were not synchronised and so are combined during the single-pulse processing.
Observations recorded in 2019 March may be affected by a buffer overflow error where the hardware was unable to keep up with the incoming data leading to either data packets being repeated or lost in the output data stream.
This may result in very small spurious signals appearing, though the total packet loss is only around 4 parts in $10^9$.
Nevertheless, to preserve data integrity, the recorded bandwidth was reduced to 642\,MHz from 2019 April until 2020 February when we had confirmed that the issue was not present in the production version of the PTUSE hardware installed in 2019 December.
The band configurations used at different times are listed in Table \ref{bands_table}.

\subsection{The Thousand Pulsar Array Single-pulse dataset}
Single-pulse data are produced for essentially all TPA observations.
In this paper we focus on the set of  observations determined by \citet{Song2023}, which lists the ``best'' single observation of each of 1192 pulsars.
This is typically the longest observation, containing slightly more than 1024 consecutive pulses from the pulsar, and observed with close to the full MeerKAT array (i.e. no sub arrays).
Around 600 of these pulsars also have much shorter follow-up observations, unusually made with a sub-array, with the aim to do timing and monitoring (see e.g. \citealp{keith24}).
We also collect single-pulse data from these follow-up sub-array observations, but due to disk space constraints we are unable to permanently archive the raw data and only the reduced data products are kept.
In order to keep the scope manageable, in this paper we discuss only the analysis of the 1192 ``best'' full-array observations.

\section{The Meertime Single-pulse Pipeline}
\label{sec:MTSP}

The MTSP was developed to process single pulse data from the PTUSE backend, primarily the large volumes of TPA single pulse data.
The MTSP is written using python, and is designed to be run over a slurm cluster environment such as that on the OzStar supercomputer\footnote{Meertime data is housed and processed on the OzSTAR supercomputer at Swinburne University of Technology.}. The pipeline performs a number of operations:
\begin{itemize}
    \item Apply polarization calibration and other corrections needed for early science data.
    \item Detection and excision of narrow-band radio frequency interference (RFI),
    \item Identify on and off-pulse phases,
    \item Split the time-series into individual pulses at a suitable off-pulse phase,
    \item Reduce the number of time and frequency samples through averaging,
    \item Generate diagnostic plots for quick visualisation of the results.
\end{itemize}

In the default configuration the output data have 1024 time samples per period, and 16 frequency channels.
However, it is possible to adjust these settings to increase the time or frequency resolution if required for a specific science case.

\subsection{Polarization Calibration}
For observations made with the L-band receiver recorded since 2020 April 11, polarization calibration is performed upstream before the data reach the PTUSE hardware, and so the data arrive at MTSP essentially fully calibrated.
For data recorded before 2020 April 11, a polarization calibration is applied in the form of a per-channel Jones matrix supplied by the South African Radio Astronomy Observatory (SARAO) for each set of pulsar observations.
Additionally, in order to form Stokes parameters, a correction for the receiver orientation is applied after calibration.
Users of the raw time-series output of the PTUSE should note that the data may have an incorrect \textsc{psrfits} header parameter for the receiver orientation, specifically that \texttt{FD\_HAND} should be set to $-1$. This is corrected automatically by the MTSP and it does not affect any downstream data products.
See \citet{2010PASA...27..104V} for a detailed explanation of the polarization conventions used by \textsc{psrfits} and \textsc{psrchive}.

\subsection{RFI Excision and on and off pulse windows}
\begin{figure}
    \centering
    \includegraphics[width=\columnwidth]{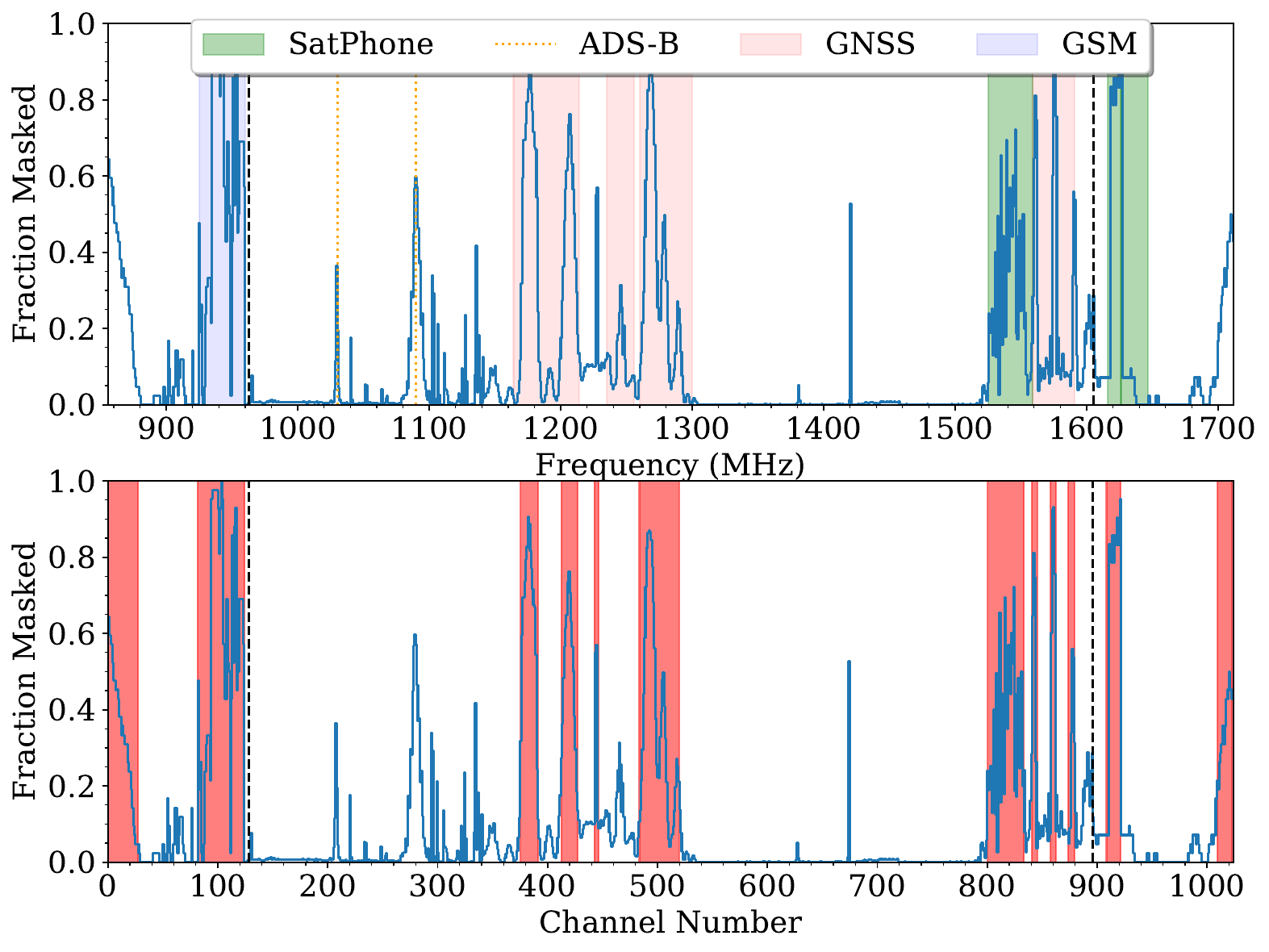}
    \caption{\label{mask}Fraction of data masked by the adaptive masking for each channel in observations during 2019 and 2020. Upper panel shows expected frequency bands for potential sources of interference. Lower panel shows the fixed channel mask applied to all data. Vertical dashed lines show the extent of data recorded in the 768-channel mode.}
\end{figure}
RFI is inherent to ground-based observations. The MTSP focuses exclusively on frequency-domain RFI mitigation by masking affected channels. We employ a fixed channel mask to target known, persistent interference, and an adaptive masking process based on off-pulse statistics. To ensure consistency for studies of pulse-to-pulse variability, the same frequency channels are removed from every pulse within a given observation.

The fixed mask (shown in the lower panel of Figure \ref{mask}) serves to increase consistency across the survey and prevents the statistics used in adaptive masking from being skewed by the most severely affected channels. These masked channels were selected based on the fraction of data masked in historical observations, and by association with known persistent RFI sources.

To adaptively identify RFI and define the on- and off-pulse regions, the time-series data are folded with \textsc{dspsr} using the pulsar ephemeris and then fully averaged in time. A coarse first-stage cleaning is performed by fitting a noise-free template\footnote{See \citet{tpaVI} and \citet{keith24} for a detailed description of template creation.} to each frequency channel using the standard FFT-convolution method implemented in \textsc{psrchive} \citep[e.g.][]{taylor92}. We define a test statistic for each channel based on the ratio of the resulting phase error to the S/N estimate; any channels where this statistic falls more than 1.5 times the interquartile range (IQR) from the median are excluded from the initial off-pulse determination.

The data are then frequency-averaged, and the template is re-fit to the profile to refine the on- and off-pulse regions. In the second cleaning stage, we calculate the off-pulse rms for every frequency channel (including those flagged in the first stage). A final mask is then applied to channels with an off-pulse rms exceeding 1.5 times the IQR. Following this final cleaning, the data are frequency-averaged once more to compute the definitive on- and off-pulse regions used for all later analysis.

\subsection{Sub-band Combination}
The time-series data recorded by PTUSE during 2019 March and 2019 April are recorded as two sub-bands split across two \textsc{psrfits} files, with a small but arbitrary start-time offset between the two band halves.
In order to simplify later work, MTSP makes an effort to produce band-combined outputs that are largely identical to the outputs from later data.
As the time-series are not exactly aligned in time, MTSP first forms de-dispersed (but not frequency-averaged) single-pulses from each sub-band, before combining them.
In this way, the time-bins are defined relative to the pulsar phase rather than the start of the observation, and so combination can be performed.
The outputs from these combined observations can be identified as the band name will be set to either ``1070+1498'' or ``1123.5+1444.5'' in the MTSP metadata.
Although all efforts are made to ensure these data are identical to those produced from later observations, we caution that there may be subtle differences in these data due to the different processing steps.

\subsection{Creation of single pulse data}
In order to create the single-pulse data, the channelised time-series data are again processed by \textsc{dspsr} in its single pulse mode.
The input ephemeris is modified to set the reference epoch (\texttt{TZRMJD}) such that 
\textsc{dspsr} splits the pulses
close to the centre of the largest off-pulse window.
Once single pulses have been written, subband combination is performed (if required), polarization calibration is applied (if required), the data are corrected for dispersion and Faraday rotation, and converted to Stokes parameters.
For TPA observations, the dispersion measure and rotation measure used are derived from the fold-mode TPA data~\citep{tpaIX}.
Finally, the pulses are averaged to 16 frequency channels, before being appended to an output \textsc{psrfits} file.
Output data are split into files containing 256 consecutive pulses in order to reduce complexity of further data handling.
A fully frequency-averaged version of the data are also created.
The output data are produced as \textsc{psrfits} data files.

\subsection{Data visualisation}
In addition to the \textsc{psrfits} data products, the pipeline also produces summary plots for each observation.
These are intended as a quick-look tool, and do not represent the full breadth of analysis that is possible with the data.
If required, the data are first averaged in phase such that that the average signal-to-noise ratio per on-pulse phase bin per period is at least 3.
In order to improve the quality of the plots, the pipeline also tries to remove variations in the baseline over the observation,
by subtracting a smooth model derived from a Gaussian process fit to the off-pulse data.
These phase-averaged and baseline-subtracted data are also exported in \textsc{numpy} \citep{numpy} ``\texttt{.npz}'' format in order to provide a quick and convenient dataset without the need the specialist code needed interpret psrfits data.
Summary plots over the whole dataset are:
\begin{itemize}
    \item Histogram of phase-averaged intensity in the on- and off-pulse regions
    \item Phase-resolved histogram of pulse intensity
    \item Phase-resolved histogram of polarization position angle.
    \item Longitude-resolved fluctuation spectrum, i.e. the Fourier transform of each pulse phase bin.
\end{itemize}
And for each stack of 256 pulses, as a function of phase and pulse number:
\begin{itemize}
    \item Stokes parameters: $I$, $Q$, $U$ and $V$.
    \item Polarization position angle
    \item Fractional linear and circular polarization
\end{itemize}

\begin{figure}
    \centering
    \includegraphics[width=\columnwidth]{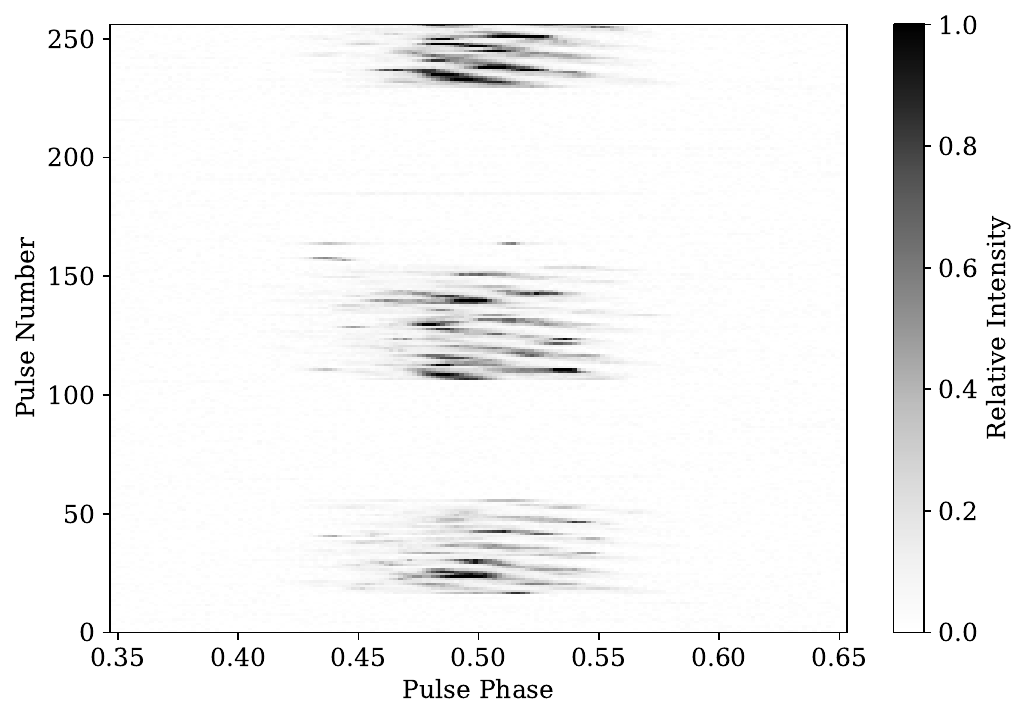}\\
    \includegraphics[width=\columnwidth]{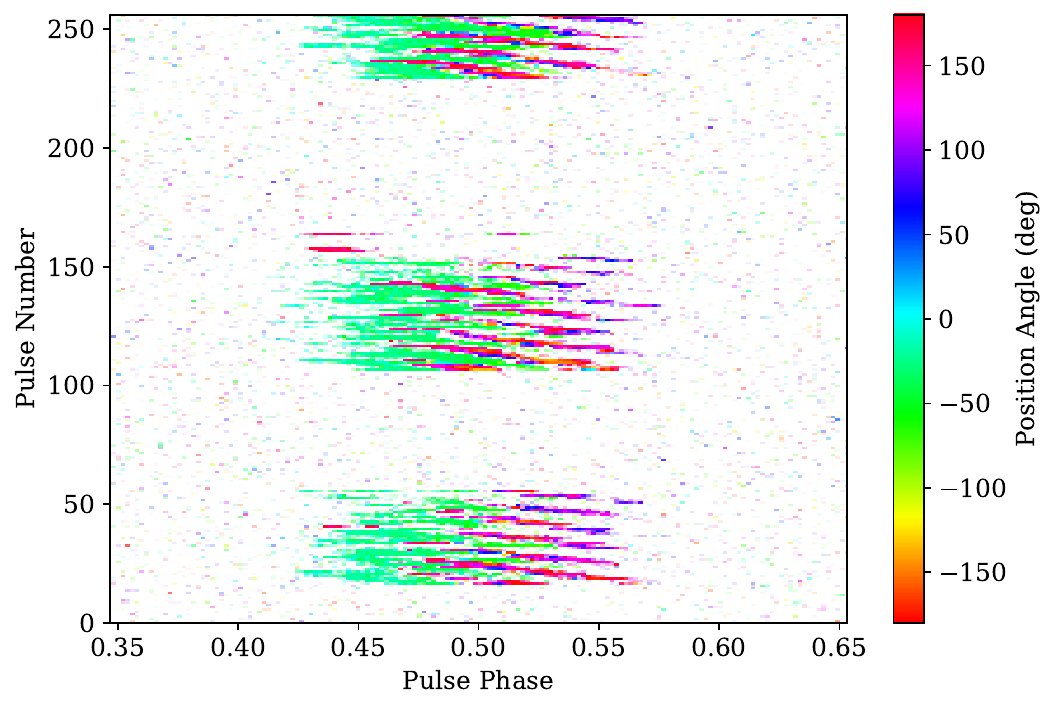}
    \caption{\label{example_plots1} Example summary plots for PSR J0034$-$0721 for a sequence of 256 pulses. Upper panel shows total intensity, lower panel shows polarization position angle.}
\end{figure}

Some example plots are shown in Figure \ref{example_plots1} for PSR J0034$-$0721, which is well known to have orthogonal polarization modes associated with its drifting sub-pulses \citep{mth73,iwjc20}.

\section{Modelling Pulse Energy Distributions}
\label{sec:energydistribs}

Phase-averaged pulse energy distributions give a measure of the intensity distribution of individual pulses. This is useful for the broad understanding of the pulse-to-pulse behaviour of the pulsar, but also can be used to compute statistics that give an indication of the potential of the dataset for further study of the pulsar behaviour.

\subsection{Data Preparation}
\begin{figure}
    \centering
    \includegraphics[width=\columnwidth]{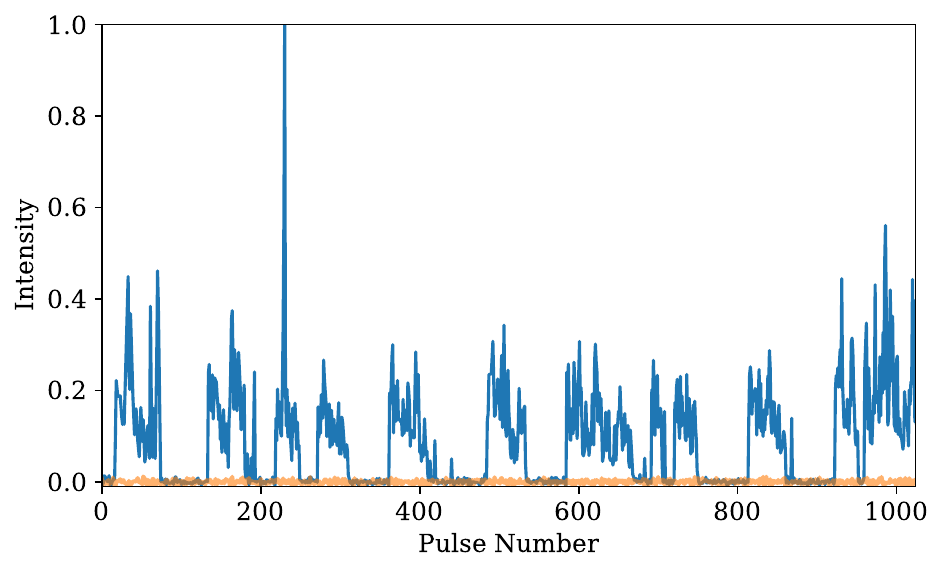}
    \caption{\label{data} The phase-averaged intensity time-series for PSR J0034$-$0721 (thin blue line). The thick orange line shows the off-pulse values for comparison.}
\end{figure}

To study the phase-averaged pulse energies, we take the frequency-averaged and baseline-subtracted output of the MTSP (the \texttt{.npz} files) and compute the pulse intensity time series over the on-pulse regions.
For comparison, we also take the average intensity in the off-pulse region of equal size.
In the case of 36 pulsars where the on-pulse region is larger than the entire off-pulse region, the off-pulse statistics are estimated by the rms noise in the off-pulse time-averaged (i.e. folded) data products and scaled to estimate the expected frequency-averaged single-pulse rms.
Time-domain RFI is excised by removing individual pulses where the rms of the entire off-pulse region falls outside 1.5 times the interquartile range.
An example on-pulse time-series is shown in Figure \ref{data} for PSR J0034$-$0721, a pulsar which exhibits both nulls and very bright individual pulses, both of which can be seen in the figure.

\subsection{Model fitting}
In order to characterise the observed pulse energies, we fit a selection of models to the data. Four `simple' single-component models are used, based on commonly observed pulse energy distributions \citep{2003MNRAS.343..512C,bjb+12}:
\begin{enumerate}
    \item \textit{Log-Normal} -- The intrinsic pulse energies are drawn from a log-normal distribution with parameters $\mu_L$, the mean of the logarithm of the data, and $\sigma_L^2$, the variance of the logarithm of the data.
    \item \textit{Normal} -- The intrinsic pulse energies are drawn from a normal distribution with mean $\mu_N$ and variance $\sigma_N^2$. In order to avoid significant negative pulses, the prior is restricted such that $\mu_N > 2.5\sigma_N$.
    \item \textit{Power-law} -- The intrinsic pulse energies are drawn from a power-law with a probability density function given by
    \begin{equation}
        f(x) = (\alpha-1)x_\mathrm{min}^{\alpha-1} x^{-\alpha},
    \end{equation}
    for $x \geq x_\mathrm{min}$, and $f(x) = 0$ otherwise, where $\alpha$ is the power-law index and $x_\mathrm{min}$ is the low-energy cut-off.
    \item \textit{Delta-function} -- The intrinsic pulse energies are of a constant value $x_0$.
\end{enumerate}
The delta-function model essentially sweeps up the cases where we have insufficient signal to determine properties of the underlying distribution.

Additionally, we fit for a number of `complex' 2 and 3 component models:
\begin{enumerate}
\setcounter{enumi}{4}
\item \textit{Log-Normal + Log-Normal},
\item \textit{Log-Normal + Normal},
\item \textit{Normal + Normal},
\item \textit{Power-law + Log-Normal},
\item \textit{Power-law + Normal},
\item \textit{Log-Normal + Log-Normal + Log-Normal},
\item \textit{Log-Normal + Log-Normal + Power-Law},
\end{enumerate}
where the + symbol indicates that pulses are drawn from one of the underlying distributions at random.
These models allow the modelling to fit more complex pulse energy distributions, including those where there are multiple pulsar emission modes.
The power-law components are required for cases where pulsars occasionally emit very bright pulses.
All models also include a nulling fraction, which we define as the probability that a pulse is drawn from the off-pulse distribution instead of the on-pulse distribution. 

The model pulse energy distribution, $f(E)$ is then given by the sum of the off-pulse distribution and the convolution of the off-pulse distribution with the intrinsic pulse energy distribution, weighted by the nulling fraction $N_f$. We can write this as
\begin{equation}
    f(E) = N_f f_\mathrm{off}(E)+(1-N_f\!)\!\int \!f_\mathrm{int}(\epsilon) f_\mathrm{off}(E-\epsilon) \mathrm{d}\epsilon ,
\end{equation}
where $f_\mathrm{int}(E)$ is the pulse energy distribution of the non-null pulses (one of (i)--(xi) above) and $f_\mathrm{off}(E)$ is the off pulse distribution.
T

We solve for the model parameters, that is the parameters for each of the probability distributions and the nulling fraction, under a Bayesian framework.
We compute the likelihood that the on-pulse data are drawn from $f(E)$, and the off-pulse data drawn from $f_\mathrm{off}(E)$
That is, for $N$ on and off-pulse measurements, the log-likelihood is given by
\begin{equation}
    \ln \mathcal{L} = \sum\limits_{i}^N \ln f(E_{\mathrm{on},i}) + \sum\limits_{i}^N \ln f_\mathrm{off}(E_{\mathrm{off},i}).
\end{equation}

To explore the parameter space, we use the Dynesty nested sampler \citep{dynesty}.
Unlike traditional Markov Chain Monte Carlo (MCMC) methods that map the posterior density directly, nested sampling is specifically designed for evidence calculation by systematically slicing the likelihood space into nested shells of known prior volume. This allows us to simultaneously estimate the posterior distribution of the parameters and reliably compute the Bayesian evidence, $Z$ for the model. In the case where several models fit the data well, we select the model with the highest $\ln Z$ for the purpose of further study of the model parameters.

The overall approach is independently developed, but largely equivalent to the Gaussian mixture model presented by \citet{ksfv18} and the Bayesian nulling estimation method of \citet{brook26}, and we would refer readers to those papers for discussion of the efficacy of the method. The primary difference in our model is that we select from a wide range of underlying pulse energy distributions rather than just a normal or log-normal distribution.
A complementary approach to null-fraction estimation in low-S/N observations has recently been presented by \citet{Grover2026}. Their method constructs distributions of randomly summed pulse energies, which statistically separate weak burst emission from the noise floor and allow null fractions to be inferred through Bayesian mixture modelling.

\section{Results and Discussion}
\label{sec:resultsdiscussion}

\subsection{Single pulse detectability}

\begin{figure}
    \centering
    \includegraphics[width=\columnwidth]{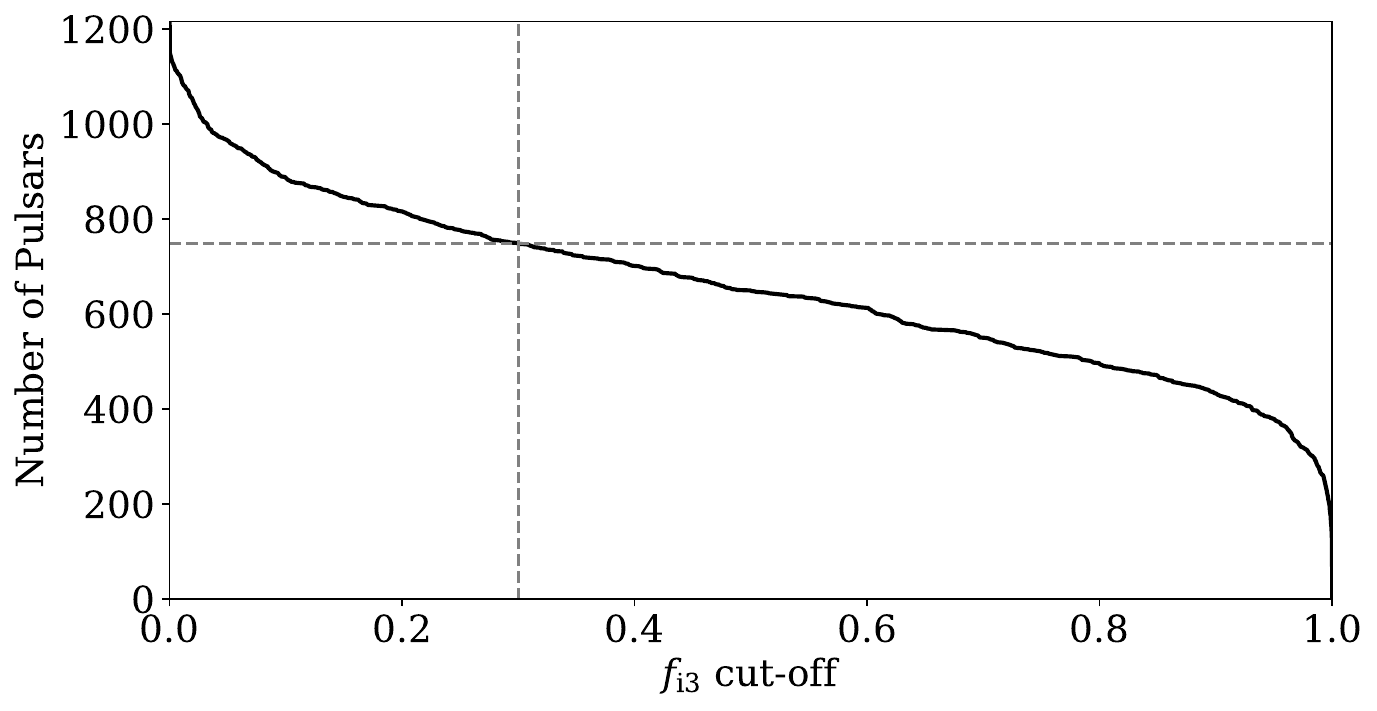}
    \caption{\label{int3s}Number of pulsars above a given cut-off in $f_\mathrm{i3}$, the fraction of the best-fit pulse energy distribution above 3-$\sigma$ from the off-pulse mean. The dashed line marks the cut-off of $f_\mathrm{i3}>0.3$ used in this paper.}

\end{figure}
Single pulse signal-to-noise ratio varies from pulse to pulse, therefore we use two metrics to quantify the fraction of detected single pulses, $f_{3\sigma}$, the fraction of pulses with phase-average intensity more than 3-$\sigma$ above that computed from the off-pulse, and $f_\mathrm{i3}$, the fraction of the best-fit intrinsic on-pulse pulse energy distribution which is more than 3-$\sigma$ above the off-pulse mean. The values for $f_{3\sigma}$ will be lower as this includes nulls, whereas $f_\mathrm{i3}$ only considers non-null pulses, but may be biased by inaccurate modelling of the underlying pulse energy distribution. In the sample, 367 pulsars have $f_\mathrm{i3}> 0.95$, i.e. 95\% of their pulses are intrinsically above 3-$\sigma$, and 260 have $f_{3\sigma} > 0.95$.  Figure \ref{int3s} shows the number of pulsars above any given $f_\mathrm{i3}$ level in our sample.

At low values of $f_\mathrm{i3}$ the pulse energy distribution modelling becomes increasingly ambiguous and small differences in the on and off pulse noise levels can adversely affect the analysis leading to, for example, false detections of nulling. 

To avoid the most ambiguous results, we applied a sample cut of $f_\mathrm{i3} > 0.3$.
This value was chosen by inspection of individual low-S/N pulsars and via simulation.
Simulations were performed by taking best-fit pulse energy distributions of a small sample high S/N pulsars and scaling them to lower flux density in order to investigate where the identification of nulling properties became unreliable.
In principle, methods such as those in \citet{Grover2026} have been demonstrated to better quantify nulling in the low S/N regime, and whilst we find it outside the scope of this paper we certainly encourage use of these methods in future work. In practice, we find that $f_\mathrm{i3} > 0.3$ produces reliable results, and therefore the remainder of the analysis will focus on the 749 pulsars for which $f_\mathrm{i3} > 0.3$.

\subsection{Nulling Fraction}

\begin{figure}
    \centering
    \includegraphics[width=\columnwidth]{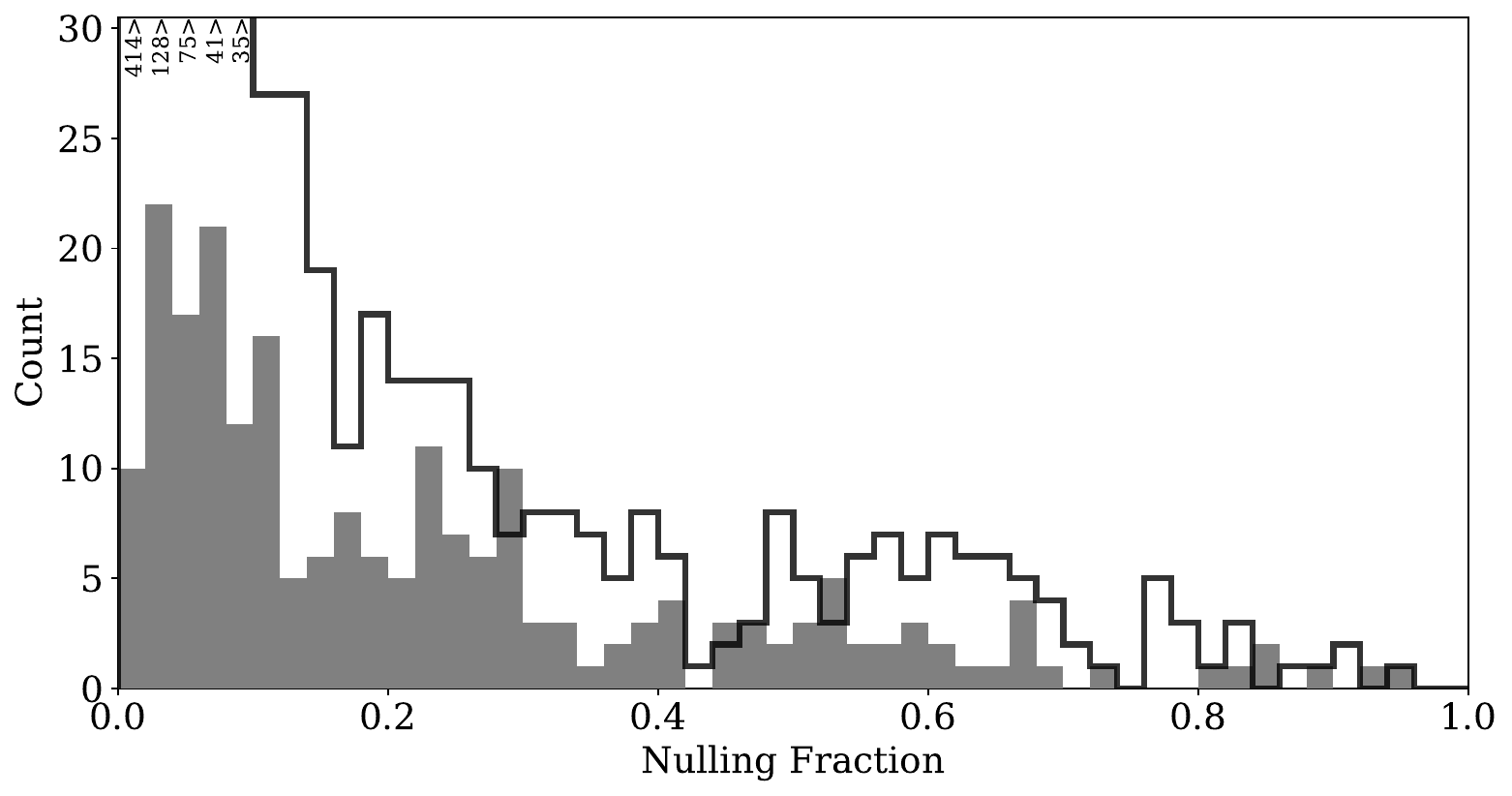}
    \caption{\label{nullfigure}Histogram of nulling fraction ($N_f$) for the sample. Solid histogram is for pulsars which show evidence for nulling, unfilled histogram is for upper-limits.}

\end{figure}

\begin{figure}
    \centering
    \includegraphics[width=\columnwidth]{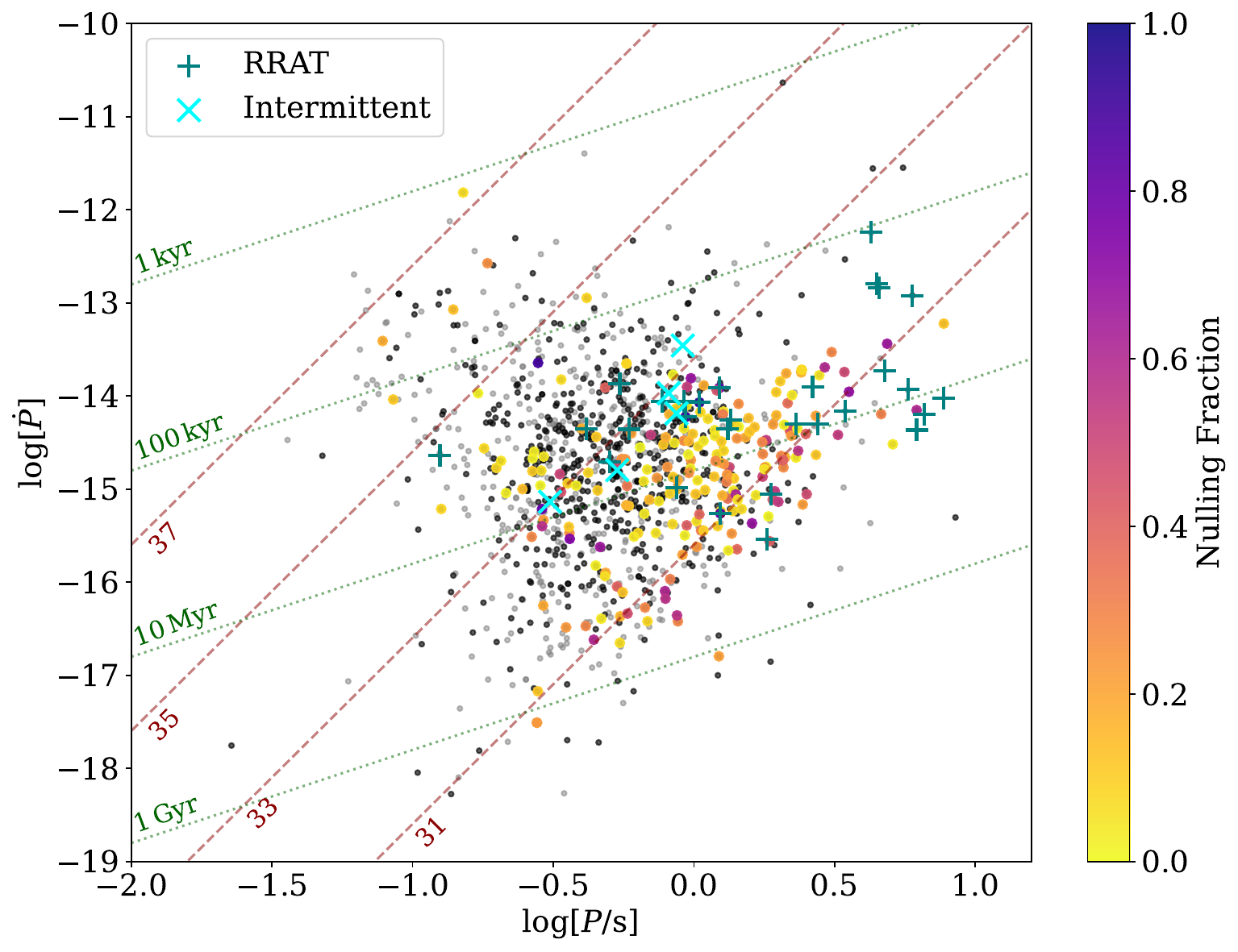}
    \caption{\label{null_ppdot}
    Nulling fraction as a function of $P$ and $\dot{P}$ for the TPA single-pulse census.
    Lines of constant $\log\!{[\dot{E}/(\mathrm{erg\,s}^{-1})]}$ and characteristic age are shown with dashed and dotted lines respectively. Black dots are pulsar for which $f_\mathrm{i3} > 0.3$, and gray dots are for all other TPA pulsars. Known RRATs and long-term nulling pulsars are overlaid with plus and cross symbols respectively.}
\end{figure}

\begin{figure}
    \centering
    \includegraphics[width=\columnwidth]{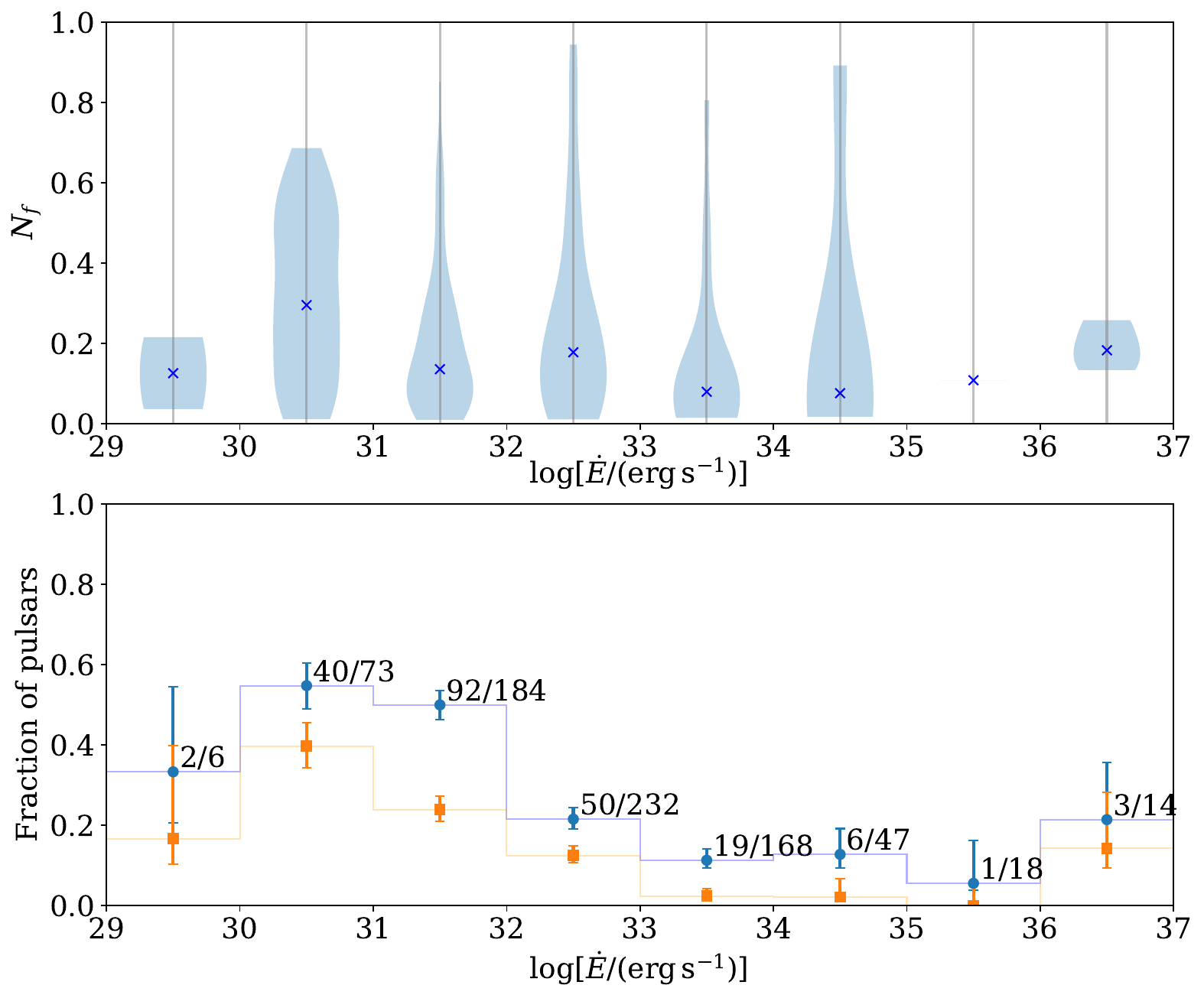}
    \caption{\label{null_vs_edot} Nulling as a function of $\dot{E}$. Upper panel, the distribution of $N_f$ for each $\dot{E}$ bin. Points indicate the median $N_f$. Lower panel, the fraction of pulsars exhibiting nulling. Circles show pulsars with $N_f>0.01$ and squares show pulsars with $N_f>0.15$. Error bars show $68\%$ bounds on the fraction of pulsars. Text values indicate the actual number of pulsars in each bin.}
\end{figure}
Figure \ref{nullfigure} shows a histogram of the nulling fraction where detected at a $>3$-$\sigma$ level, as well as upper limits for all other pulsars. These results are also shown as a function of $P$ and $\dot{P}$ on Figure \ref{null_ppdot}.
There is a clear indication that the nulling fraction and number of nulling pulsars evolves across the {\ppdot} diagram.
The nulling pulsars are mainly below  $\dot{E}\sim 10^{33}\,\mathrm{erg\,s}^{-1}$ with a band of high  null fractions appearing around $\dot{E}\sim 10^{31} \,\mathrm{erg\,s}^{-1}$.
It is perhaps interesting that the `intermittent pulsars' which exhibit very long-term nulls appear at the higher end of this $\dot{E}$ range ($\sim 10^{33}\,\mathrm{erg\,s}$; \citealp{lsf+17}), and whilst there are some high null-fraction pulsars around this region of the \ppdot\ diagram there does not generally seem to be any significant difference in the properties of our sample close to this $\dot{E}$.

In Figure \ref{null_vs_edot} we show the distribution of nulling fraction, as well as the fraction of pulsars showing significant nulling as a function of $\dot{E}$. Errors on the fraction of nulling pulsars are estimated by modelling the fraction as the success probability of a binomial distribution and deriving credible intervals from the corresponding beta posterior distribution (assuming a uniform prior).
The general trend is a gradual decrease of nulling phenomena with increasing $\dot{E}$, though there is some hint that there is an excess of nulling pulsars at high $\dot{E}$.
Inspection of the 4 nulling pulsars with $\dot{E}>10^{35}\,\mathrm{erg\,s}^{-1}$ indicates they are all relatively low signal-to-noise cases with only a statistical detection of nulling rather than clear periods of zero emission.
Combined with the small numbers of pulsars in these bins, we therefore suggest that this apparent rise at high $\dot{E}$ is not significant.

To investigate the evolution of nulling fraction on the {\ppdot} diagram in more detail, we can model it as being drawn from the Beta distribution, which is highly flexible for bounded random values.

We therefore model the  $N_f$ of any given pulsar as a random parameter drawn from a Beta distribution with shape parameter $v$, having probability density 
\begin{equation}
    f(N_f;\mu,v) = k N_f^{\mu v -1}(1-N_f)^{(1-\mu)v},
\end{equation}
where $k$ is a normalisation constant, and mean $\mu(P,\dot{P})$ given by,
\begin{equation}
\mu(P,\dot{P}) = m\left(\frac{P}{\mathrm{s}}\right)^{\alpha} \left(\frac{\dot{P}}{10^{-15}}\right)^{\beta}.
\end{equation}
The parameters $\alpha$, $\beta$ therefore represent the evolution of the mean nulling fraction over $P$ and $\dot{P}$ respectively, and $m$ as the typical $N_f$ for a $P=1\,\mathrm{s}$ $\dot{P}=10^{-15}$ pulsar.

We then compute a log-likelihood summed over all pulsars $p$,
\begin{equation}
\ln \mathcal{L}(\alpha,\beta,m,v) =\sum\limits_p \ln[C_B(N_{u,p},\mu_p,v) - C_B(N_{l,p},\mu_p,v)],
\end{equation}
where $C_B$ is the cumulative distribution function of the Beta distribution, and $N_{u,p}$ and $N_{l,p}$ are $95\%$ upper and lower bounds on $N_f$ for each pulsar. This is in effect reducing the $N_f$ posterior estimates to a uniform distribution, and is done for computational simplicity.
We then maximise the likelihood, solving for $\alpha$, $\beta$, $m$, and $v$ using common MCMC techniques provided by \textsc{emcee} \citep{fhlg13}.
The results are $\alpha = 0.93 \pm 0.08$, $\beta=-0.10 \pm 0.03$, $m = 0.08 \pm 0.01$ and $v=0.19 \pm 0.05$.
This implies a nearly linear dependence of the nulling fraction with period, and little dependence on period derivative. 
The strong dependence of nulling fraction on spin period suggests that nulling is more closely linked to the large-scale structure of the magnetosphere than to available spin-down power. One possible interpretation is that nulling becomes more likely as the light-cylinder radius increases and the polar cap shrinks, potentially making the magnetosphere more susceptible to perturbations that intermittently disrupt radio emission. In this context, \cite{Wright1979} and later \citet{cordes2008} proposed that episodic accretion of circumstellar debris could trigger changes in magnetospheric state, leading to nulls or intermittent behaviour. While that model does not necessarily predict a scaling of infall rate with light-cylinder radius, the effectiveness of such external perturbations may plausibly depend on the size and geometry of the magnetosphere, which evolve strongly with spin period.

However, this is certainly not a unique explanation of the observed nulling behaviour, and these results should be considered over a wide range of theoretical models for pulsar nulling.
Further, our analysis does not consider the timescale of the nulling, which also is seen to evolve with $P$ and $\dot{P}$ \citep{lsf+17}, as does the apparent quasi-periodicity of null sequences in some pulsars \citep{Grover2026}, and there is evidence for decade-timescale evolution of nulling properties in some pulsars \citep{brook26}.

\subsection{Pulse Energy Distributions}

\begin{figure*}
    \centering
    \begin{tabular}{c|c}
    \includegraphics[width=0.5\linewidth]{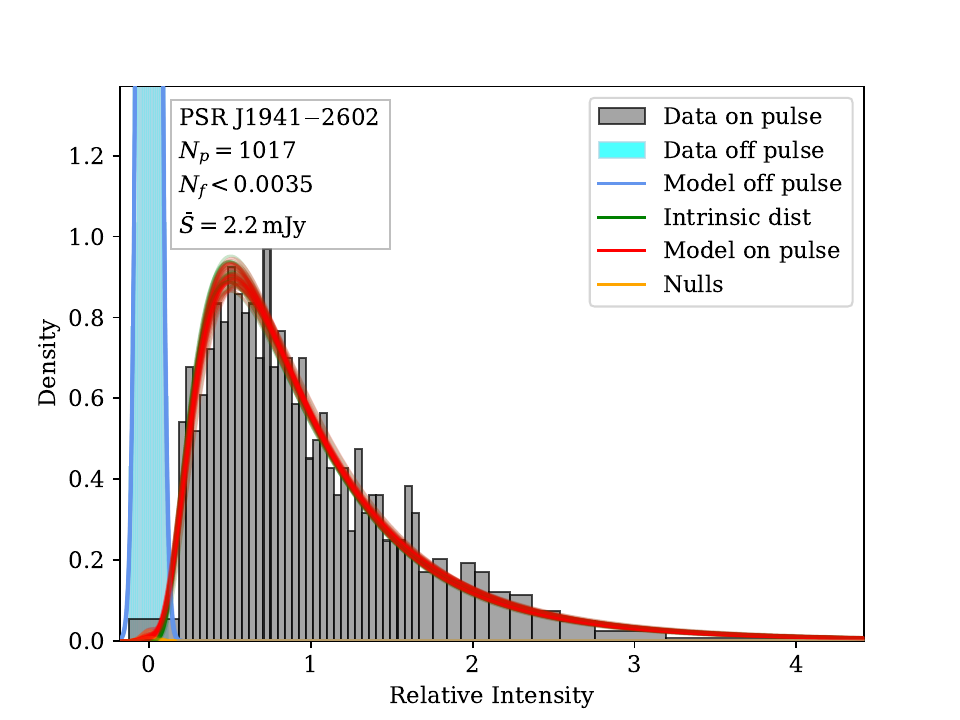} &
         \includegraphics[width=0.5\linewidth]{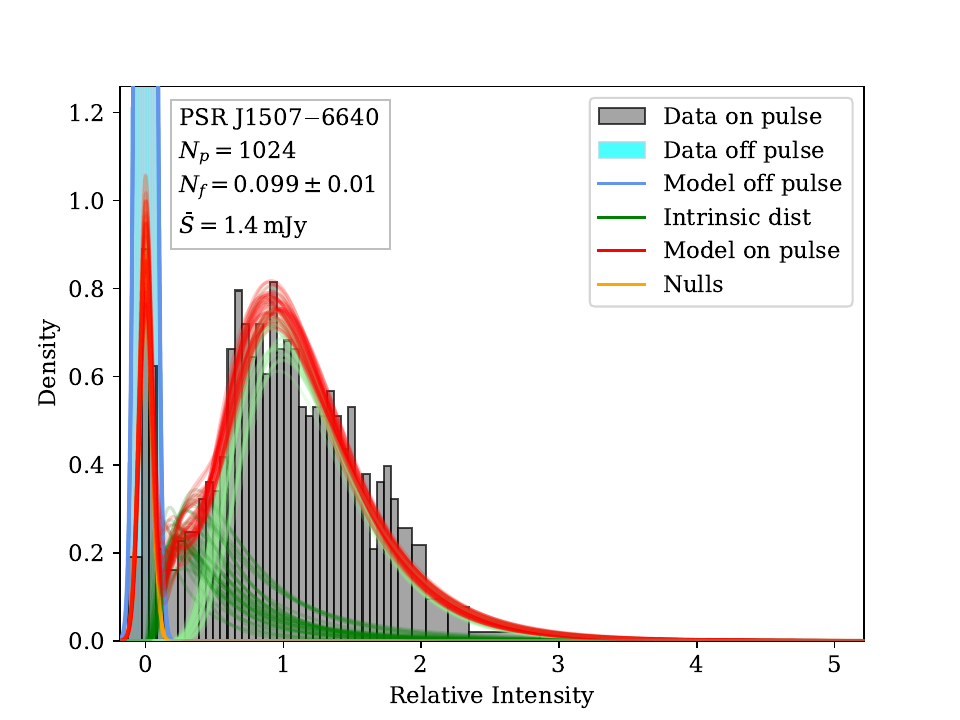} \ \\
     
    \end{tabular}
    
    \caption{Visualisations of the pulse energy distribution of two example pulsars. Curves show samples from the posterior distribution for the preferred pulse energy distribution model, and histograms show the observed on and off-pulse intensities. Intensity is normalised such pulses with the mean flux density have unit intensity. The inset box shows the number of pulses ($N_p$), the nulling fraction ($N_f$) and the mean flux density ($\overline{S}$). Left panel shows PSR J1941$-$2602, well fit by a single log-normal component without any nulling. Right panel shows PSR J1507$-$6640, which requires at least two model components (shown here two log-normal components) and a nulling fraction of $\sim1\%$. Similar figures for all pulsars are available in the associated data release.}
    \label{example_ped}
\end{figure*}

\begin{figure}
    \centering
    \includegraphics[width=0.8\columnwidth]{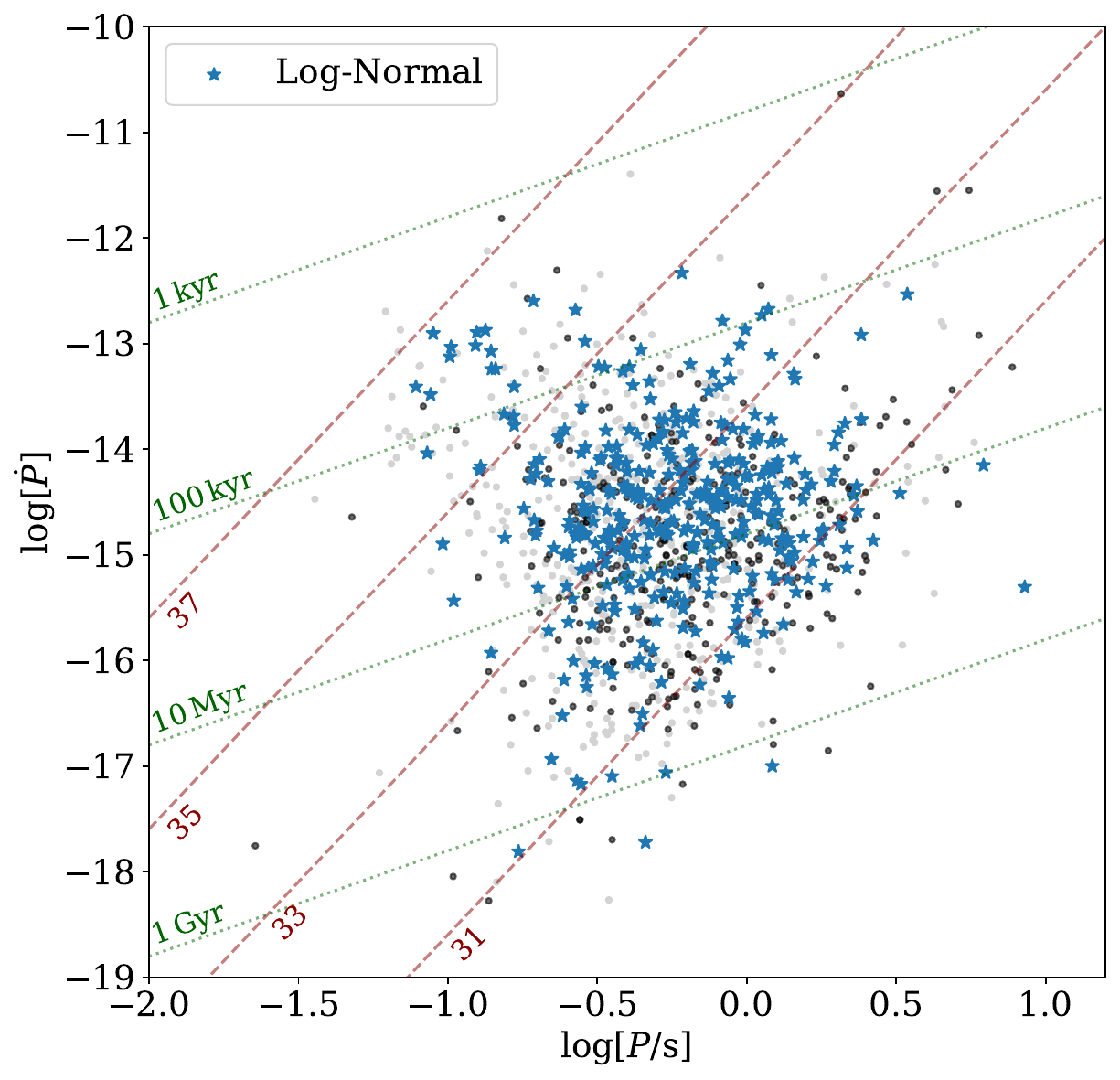}
    \caption{Distribution of pulsars requiring log-normal components in their pulse energy distribution as a function of $P$ and $\dot{P}$, shown with blue stars. Black dots are all pulsars for which $f_\mathrm{i3} > 0.3$, and gray dots are for all other TPA pulsars.  Lines of constant $\log\!{[\dot{E}/(\mathrm{erg\,s}^{-1})]}$ and characteristic age are shown with dashed and dotted lines respectively. }
    \label{ppdot_lnorm}
\end{figure}

\begin{figure}
    \centering
    \includegraphics[width=\columnwidth]{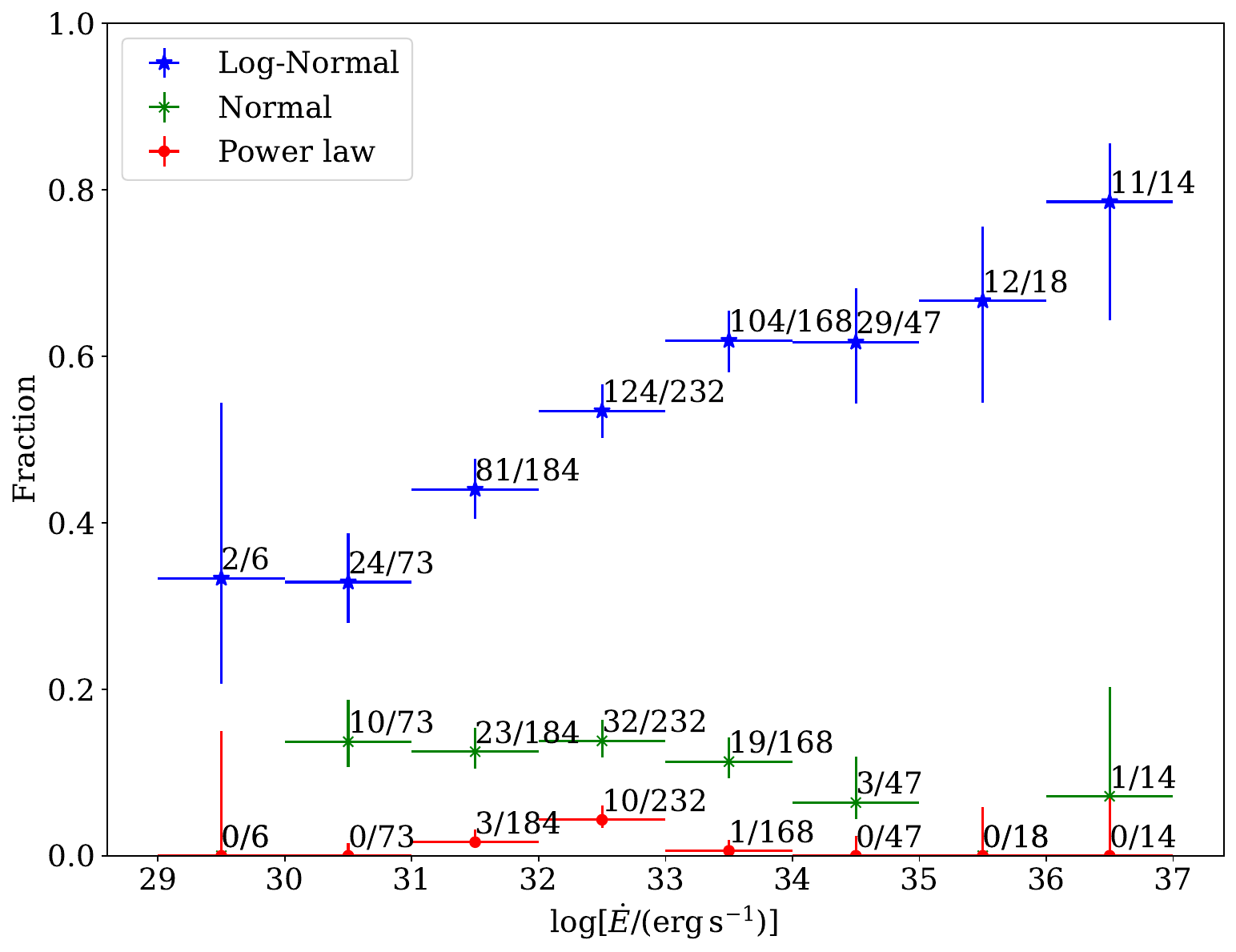}
    \caption{\label{edot_vs_type} Pulse energy distribution shape as a function of $\dot{E}$. This shows distribution shapes that are required for all best fitting models, hence pulsars can lie in multiple shape categories, or none, and hence the sum of categories is not expected to sum to the total number of pulsars. Error bars show $68\%$ intervals estimated as in Figure~\ref{null_vs_edot}. The number of pulsars in each bin is shown as a fraction next to each point.}
\end{figure}

We estimate the log evidence, $\ln Z$, for each of the pulse energy distribution models, and reject any models with $\ln Z - \ln Z_\mathrm{best} < -2.5$. 
For the majority of pulsars this Bayesian model selection indicates more than one model that is compatible with the observed pulse energy distribution.
371 (51 per cent) of the pulse energy distributions are able to be explained with a single component model and 360 (49 per cent) require at least two components.
Although the very low S/N pulsars are biased towards simple models, the fraction of pulsars requiring more than two components remains $\sim 50$ per cent even if we consider only the pulsars with $f_{3\sigma} > 0.95$.
Figure \ref{example_ped} shows two examples pulse energy distribution fits for pulsars with  $f_{3\sigma} > 0.95$. PSR J1941$-$2602 shows a log-normal distribution with effectively no nulls. PSR J1507$-$6640 shows both nulling (null fraction $\sim 1\%$) as well as an on-pulse energy distribution that requires at least two components, and the highest evidence is for the three log-normal component model.

With such a large sample of pulsars and multiple models that can fit many pulsars, we can try to simplify the dataset by classifying the pulsars based on which models are compatible with the observed pulse energy distribution.
Pulsars that are compatible with exactly one of the single-component models can be classed as log-normal (278 pulsars), normal (34 pulsars) or power-law (10 pulsars).
Of the remainder, 49 pulsars are compatible with more than one single-component model, e.g. we cannot distinguish between log-normal and normal, and 366 are only compatible with the more complex multi-component models, typically with more than one viable model. The remaining 12 pulsars are most compatible with the delta-function model and hence we are unable to discern anything about the underlying pulse energy distribution.

Given the large number of `complex' pulse energy distributions, we can also classify pulsars based on any distribution types that are required across all preferred models for a given pulsar.
For example if all preferred models require a log-normal distribution we classify this pulsar as  `needs log-normal'.
It is possible for a pulsar to require more than one type of distribution and therefore appear in multiple `needs' categories.
Figure \ref{ppdot_lnorm} shows the distribution of the pulsars requiring log-normal components across the \ppdot\ diagram (the other two categories are shown in Appendix~\ref{extra_ppdots}).
The density of log-normal pulse energy distributions appears to increase with increasing $\dot{E}$.
To investigate this further, Figure \ref{edot_vs_type} shows the number of pulsars in each `needs' category as a function of $\dot{E}$.
Although the numbers of pulsars in some bins is small, there does seem to be a subtle evolution of the pulse energy distributions, with more prevalence of log-normal components at high $\dot{E}$.
There may also be a preference for power-law components around $\dot{E}\sim10^{32}\,\mathrm{erg\,s}^{-1}$, and there is other evidence for complex emission behaviour in this $\dot{E}$ range. 
Long-term intermittent pulsars seem to lie along this $\dot{E}$ \citep{lsf+17}, and there is evidence that there is a change in the pulse profile properties around $\dot{E}\simeq10^{32.5}\,\mathrm{erg\,s}^{-1}$ \citep{corecone_edot}.
It is likely that a phase-resolved study is necessary to properly understand if there is a physical relationship here.


\subsection{Single-pulse luminosity}
The single-pulse energy distributions are initially computed relative to the mean on-pulse energy. 
Although these individual single-pulse observations are not flux-calibrated, the corresponding folded profiles (which make use of the same underlying astrophysical signal) are automatically calibrated by the MeerTime pipeline. 
We can therefore approximate the absolute single-pulse energy in flux units by scaling the relative energies by the time-averaged flux density derived from the fold pipeline \citep{tpaIX}.

We convert this single-pulse flux density to an estimate of the bolometric radio luminosity by assuming a power-law spectral energy distribution with a constant spectral index of $-1.2$ between 10\,MHz and 100\,GHz. 
This effective spectral index was selected because it minimizes the systematic divergence in integrated luminosity across the overlapping sample of 42 pulsars shared with \citet{Bilous2020}, who utilized a comprehensive spectral energy distribution model measured over a wide range of frequencies.
Following the beaming model of \citet{tpaXV}, and utilizing the empirical pulse widths determined for the TPA sample \citep{tpaVI}, the estimated bolometric radio luminosity is given by:

\begin{equation}
    L = 4\pi D^2 S_0 \frac{\sin^2{(\pi\delta_{10})}}{\delta_{10}}  \int_{10\,\mathrm{MHz}}^{100\,\mathrm{GHz}} \left(\frac{\nu}{\nu_0}\right)^{-1.2} \,\mathrm{d}\nu,
\end{equation}
where $D$ is the distance to the pulsar, $\delta_{10}$ is the pulse width at 10\% of the peak intensity expressed as a fraction of the pulse period, and $S_0$ is the mean flux density measured at the reference frequency $\nu_0 = 1.28\,\mathrm{GHz}$. 
While these values provide a useful proxy for the true radio luminosity, we caution that significant unmodeled uncertainties remain. In particular, this metric relies on a highly simplified model of the spectral energy distribution, and the large distance uncertainties inherent to many pulsars in the sample will introduce corresponding scatter into the calculated luminosities.

\begin{figure}
    \centering
    \includegraphics[width=\columnwidth]{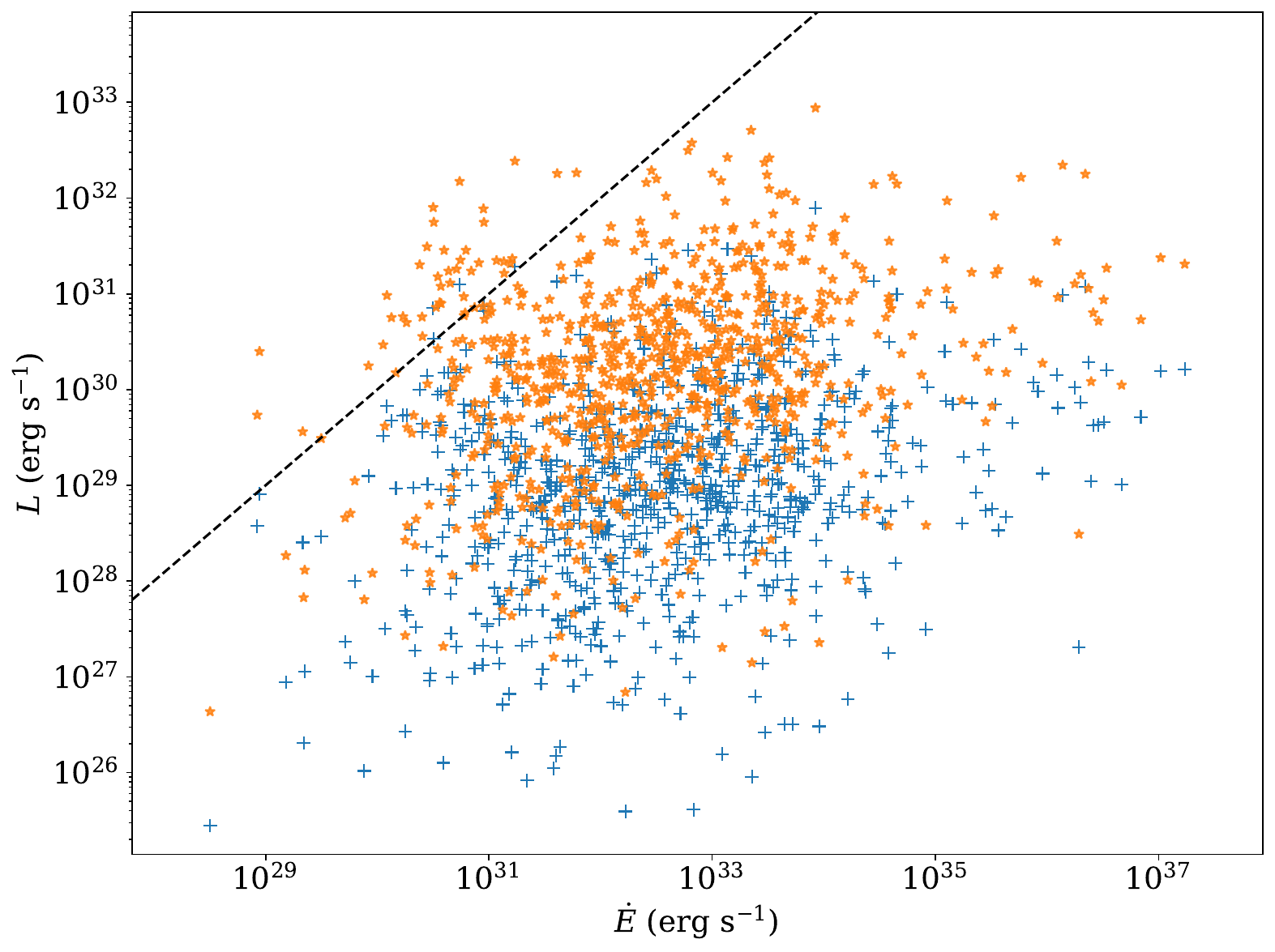}
    \caption{\label{luminosity_vs_edot.pdf} Estimated radio luminosity of the TPA sample for the average flux density (plus symbol) and for the brightest observed pulse (stars), as a function of the inferred rotational energy loss rate $\dot{E}$. Dashed line shows the one-to-one correspondence of luminosity and $\dot{E}$}
\end{figure}

Figure \ref{luminosity_vs_edot.pdf} shows $L$ as a function of $\dot{E}$ for the median flux density (i.e. that taken from the time-averaged observation) and from the brightest observed single pulse.
We can see that many pulsars, especially those below $\dot{E} = 10^{32}\,\mathrm{erg~s^{-1}}$ show individual pulses with luminosity close to or exceeding the long-term spin-down energy loss rate by 1--2 orders of magnitude, even where their time-averaged luminosity does not.

This suggests that the high luminosities ($\gtrsim\dot{E}$) observed in ultra-long period transients such as GPM~J1839$-$10 \citep{long_period_nhw} may not be unique to such objects.
Indeed, extrapolating from the TPA population, we might expect a pulsar with $P$ and $\dot{P}$ similar to GPM~J1839$-$10 would regularly show bright pulses with $L_p$ in excess of $\dot{E}$ by a factor of 10 or more.

\section{Conclusions}
\label{sec:conclusions}

In this work we have presented the Thousand Pulsar Array single-pulse data set and the MTSP pipeline used to produce this data set. The public data release of the `census' observations is avaliable at \href{http://doi.org/10.5281/zenodo.18980771}{\texttt{doi.org/10.5281/zenodo.18980771}}, and can also be accessed via a web interface at \href{http://psrweb.jb.man.ac.uk/tpa/singlepulse/}{\texttt{psrweb.jb.man.ac.uk/tpa/singlepulse}} which will be updated with future data releases in time. We calculate nulling fractions and model the single-pulse energy , which are also avaliable online at \href{http://psrweb.jb.man.ac.uk/tpa/pulse_energy/}{\texttt{psrweb.jb.man.ac.uk/tpa/pulse\_energy}} and archived on Zenodo at \href{http://doi.org/10.5281/zenodo.18982781}{\texttt{doi.org/10.5281/zenodo.18982781}}.

The scale of the TPA single-pulse data set enables studies of the population-level statistics of single-pulse energy distributions at a new level of sensitivity. The results show that single-pulse energies evolve across the pulsar population, with some evidence of evolution of both nulling fraction and energy distribution shape with $P$ or $\dot{E}$. 
However, they also show that the considerable complexity of pulsar radio emission variability is not sufficiently well captured by phase-averaged measurements.
For example, measurements of phase-integrated pulse energy distributions cannot capture the fact that pulse profiles are made up of individual sub-pulses, which have complex behaviour such as sub-pulse drifting and mode-changing. Future work should therefore model phase-resolved single pulse energies and investigate their relationships both to spin-down and to the diverse ways in which time-variability manifests in pulsar observations.

\section*{Acknowledgements}
The MeerKAT telescope is operated by the South African Radio
Astronomy Observatory (SARAO), which is a facility of the National
Research Foundation, an agency of the Department of Science
and Innovation. SARAO acknowledges the ongoing advice and
calibration of GPS systems by the National Metrology Institute
of South Africa (NMISA) and the time space reference systems
department of the Paris Observatory.
Pulsar research at Jodrell Bank is supported by a consolidated grant (ST/X001229/1) from the UK Science and Technology Facilities Council (STFC).
PTUSE was developed with
support from the Australian SKA Office and Swinburne University
of Technology. This work made use of the OzSTAR national HPC
facility at Swinburne University of Technology. MeerTime data are
housed on the OzSTAR supercomputer. The OzSTAR programme
receives funding in part from the Astronomy National Collaborative
Research Infrastructure Strategy (NCRIS) allocation provided by
the Australian Government. 
\section*{Data Availability}

The single-pulse datasets for the TPA single-pulse census used in this paper are archived via Zenodo \href{http://doi.org/10.5281/zenodo.18980771}{\texttt{doi.org/10.5281/zenodo.18980771}}, and can also be accessed via a web interface at \href{http://psrweb.jb.man.ac.uk/tpa/singlepulse/}{\texttt{psrweb.jb.man.ac.uk/tpa/singlepulse}}. 
The results of the pulse energy distribution fitting are available at \href{http://psrweb.jb.man.ac.uk/tpa/pulse_energy/}{\texttt{psrweb.jb.man.ac.uk/tpa/pulse\_energy}} archived on Zenodo \href{http://doi.org/10.5281/zenodo.18982781}{\texttt{doi.org/10.5281/zenodo.18982781}}. The software used to generate the data is available at \href{http://github.com/SixByNine/meertimesinglepulse/}{\texttt{github.com/SixByNine/meertimesinglepulse}}



\bibliographystyle{mnras}
\bibliography{mtsp_updated} 



\appendix

\section{Additional $P$-$\dot{P}$ diagrams}
\label{extra_ppdots}

\begin{figure}
    \centering
    \includegraphics[width=0.8\columnwidth]{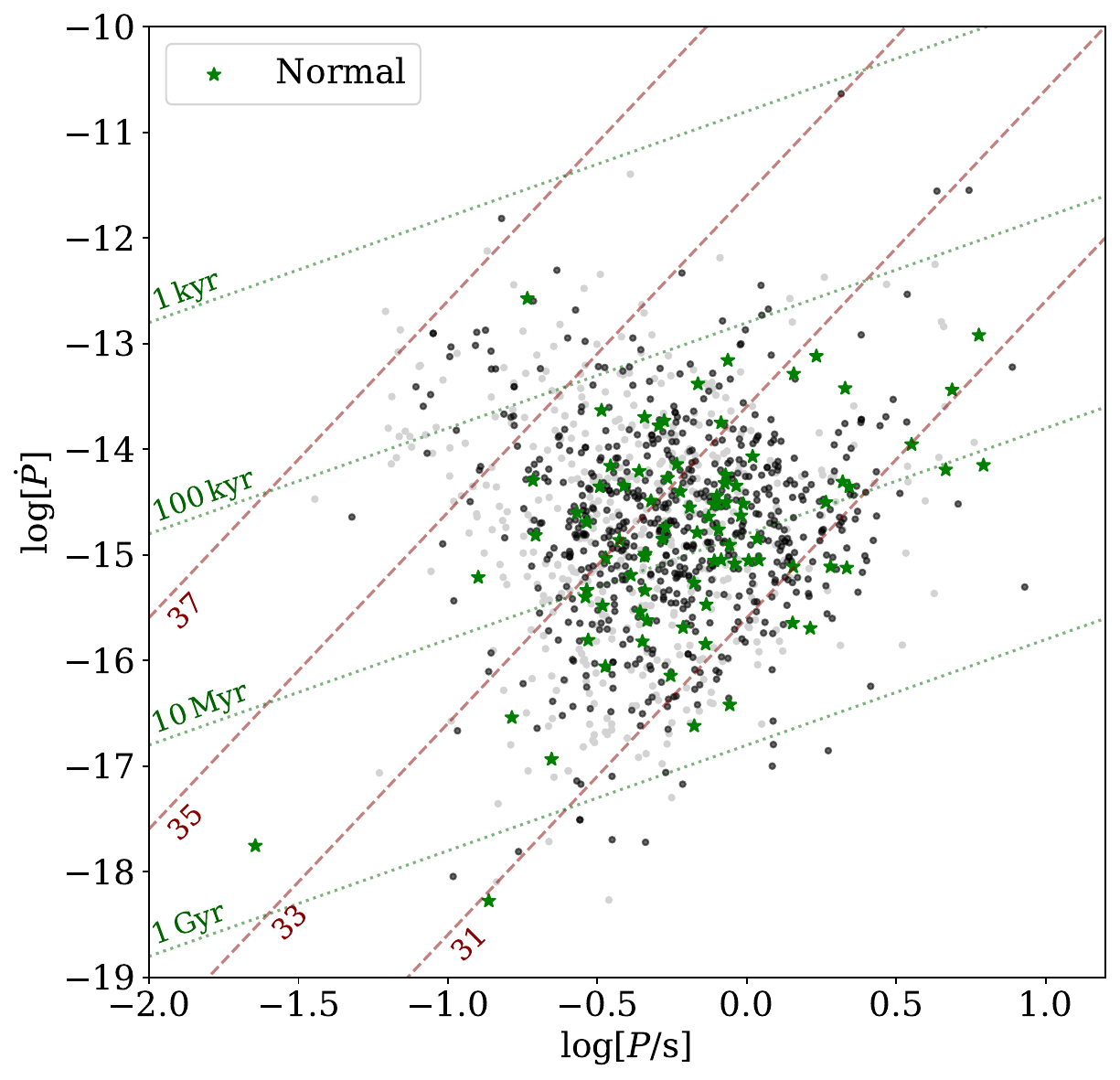}
    \caption{As Figure~\ref{ppdot_lnorm} but for pulsars requiring power-law components.}
    \label{ppdot_norm}
\end{figure}

\begin{figure}
    \centering
    \includegraphics[width=0.8\columnwidth]{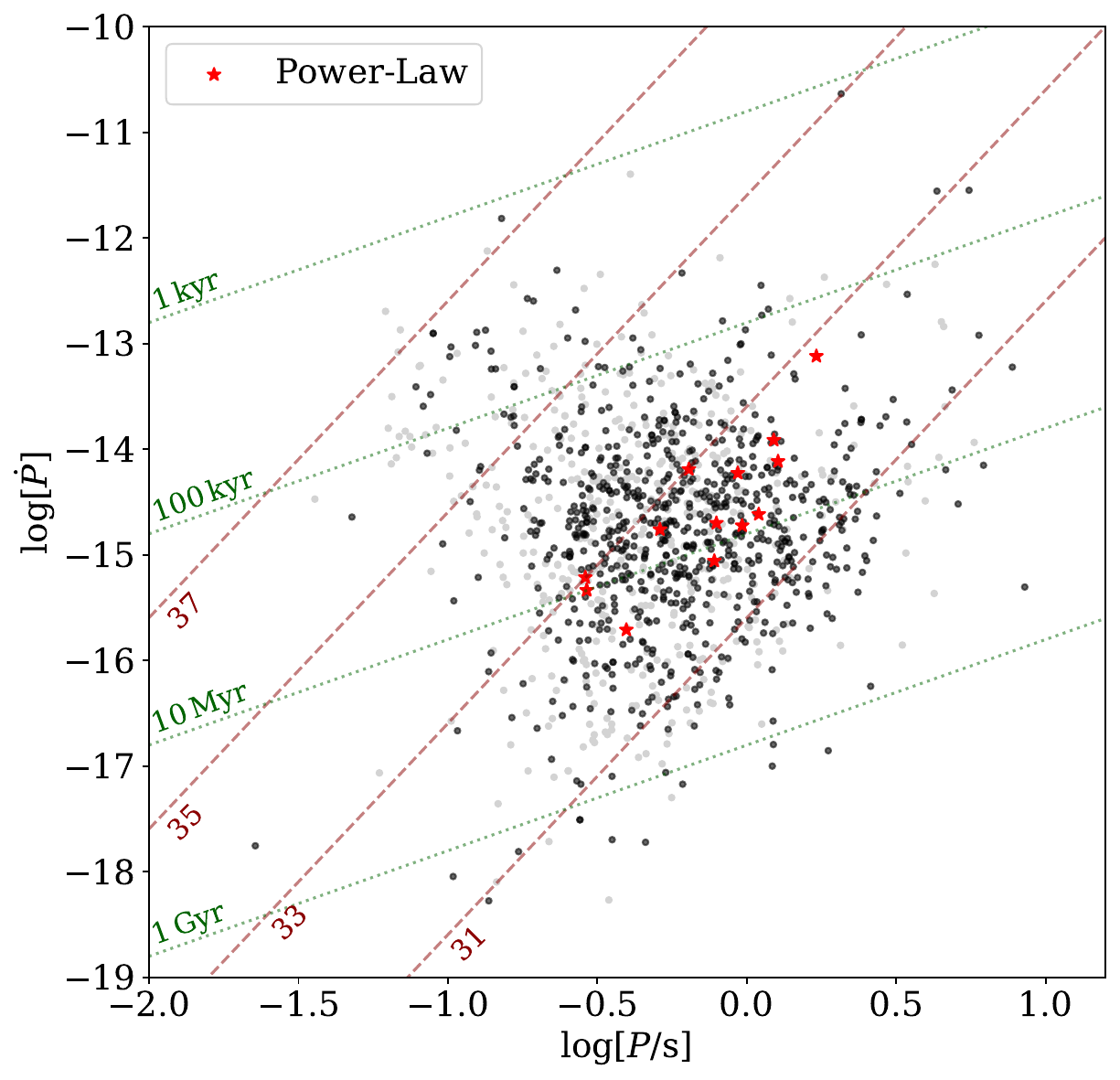}
    \caption{As Figure~\ref{ppdot_lnorm} but for pulsars requiring power-law components.}
    \label{ppdot_pl}
\end{figure}

Here we show the distribution of the pulsars which require normal (Figure \ref{ppdot_norm}) and power-law components (Figure \ref{ppdot_pl}) across the \ppdot\ diagram.


\bsp	
\label{lastpage}
\end{document}